\documentclass[twocolumn,showpacs,preprintnumbers,amssymb,pra,%
floatfix,superscriptaddress,a4paper]{revtex4}

\usepackage{etoolbox}
\usepackage{color}
\usepackage{placeins}
\usepackage{mathrsfs}
\usepackage{amsmath}
\usepackage{graphicx}
\usepackage{graphicx,color}
\usepackage{dcolumn}
\usepackage[nice]{nicefrac}
\usepackage[tight]{units}

\usepackage{MnSymbol}
\def\t0{\mbox{$t_{\mbox{{\tiny {0}}}}$}}
\def\p0{\mbox{$p_{\mbox{{\tiny {0}}}}$}}
\def\E0{\mbox{$E_{\mbox{{\tiny {0}}}}$}}

\newcommand{\be}{\begin{equation}}
\newcommand{\ee}{\end{equation}}
\newcommand{\ba}{\begin{eqnarray}}
\newcommand{\ea}{\end{eqnarray}}

\def\w{$\omega$}
\def\w2w{$\omega/2\omega$}

\newcommand{\lund}{Department of Physics, Lund University, P. O. Box 118, SE-22100 Lund, Sweden}
\newcommand{\lundSynch}{Department of Synchrotron Radiation Instrumentation, Max IV laboratory, Lund University, P.O. Box 118, SE-221 00 Lund, Sweden}
\newcommand{\albanova}{Department of Physics, Stockholm University, AlbaNova University Center, SE-106 91 Stockholm, Sweden}
\newcommand{\uam}{Departamento de Qu\'imica, M\'odulo 13, Universidad Aut\'onoma de Madrid, 28049 Madrid, Spain}
\newcommand{\imdea}{Instituto Madrile\~no de Estudios Avanzados en Nanociencia (IMDEA-Nanociencia), Cantoblanco, 28049 Madrid, Spain}
\newcommand{\ifimac}{Condensed Matter Physics Center (IFIMAC), Universidad Aut\'onoma de Madrid, 28049 Madrid, Spain}
\newcommand{\upmc}{Laboratoire de Chimie Physique-Mati\`ere et Rayonnement, Universit\'e Pierre et Marie Curie, 11, Rue Pierre et Marie Curie, 75231 Paris Cedex, 05, France}

\begin{document}

\title{Phase measurement of a Fano window resonance using tunable attosecond pulses}

\author{M.~Kotur}\affiliation{\lund}
\author{D.~Gu\'enot}\affiliation{\lund}
\author{\'A.~Jim\'enez-Gal\'an}\affiliation{\uam}
\author{D.~Kroon}\affiliation{\lund}
\author{E.~W. Larsen}\affiliation{\lund}
\author{M.~Louisy}\affiliation{\lund}
\author{S.~Bengtsson}\affiliation{\lund}
\author{M.~Miranda}\affiliation{\lund}
\author{J.~Mauritsson}\affiliation{\lund}
\author{C.~L.~Arnold}\affiliation{\lund}
\author{S.~E.~Canton}\affiliation{\lundSynch}
\author{M.~Gisselbrecht}\affiliation{\lund}
\author{T.~Carette}\affiliation{\albanova}
\author{J.~M.~Dahlstr\"om}\affiliation{\albanova}
\author{E.~Lindroth}\affiliation{\albanova}
\author{A.~Maquet}\affiliation{\upmc}
\author{L.~Argenti}\affiliation{\uam}
\author{F.~Mart\'{\i}n}\affiliation{\uam}\affiliation{\imdea}\affiliation{\ifimac}
\author{A.~L'Huillier}\affiliation{\lund}

\begin{abstract}
We study the photoionization of argon atoms close to the 3s$^2$3p$^6$ $\rightarrow$ 3s$^1$3p$^6$4p $\leftrightarrow$ 3s$^2$3p$^5$ $\varepsilon \ell$, $\ell$=s,d Fano window resonance. An interferometric technique using an attosecond pulse train, i.e. a frequency comb in the extreme ultraviolet range, and a weak infrared probe field allows us to 
study both amplitude and phase of the photoionization probability amplitude as a function of photon energy. A theoretical calculation of the ionization process accounting for several continuum channels and bandwidth effects reproduces well the experimental observations and shows that the phase variation of the resonant two-photon amplitude depends on the interaction between the channels involved in the autoionization process.  
\end{abstract}

\maketitle
\noindent 

Electron dynamics induced by photoabsorption are of fundamental importance in nature. 
The development of table-top attosecond-duration sources in the extreme ultraviolet (XUV) spectral range has opened up possibilities of accessing these dynamics 
by coherent pump--probe experiments with attosecond resolution. 
In recent experiments, photoemission delays have been measured using attosecond pulses combined with an infrared (IR) probe field in a variety of systems, from solid state to gas samples \cite{CavalieriNature2007,LocherOptica2015,SchulzeScience2010,KlunderPRL2011,GuenotJPB2014,PalatchiJPB2014,ManssonNP2014}.

Photoionization dynamics are strongly affected by the presence of resonances. 
When a highly excited bound state is correlated with an open ionization channel,
autoionization may occur by the ejection of an electron and relaxation of the remaining core. Interference between the direct- and the autoionizing pathways leads to the characteristic asymmetric Fano profiles in the photoionization cross-section \cite{BeutlerZP1935,FanoPR1961}, which have been extensively measured using synchrotron radiation (see e.g. argon spectra in \cite{BerrahJPB1996,SorensenPRA1994}).  

Studying and even controlling this interaction has been a major goal of attosecond science since the early days and several methods have been developed to this end \cite{KrauszRMP2009}. Attosecond streaking was used for the first time-resolved measurement of Auger decay in Kr \cite{DrescherNature2002}, while the more recent transient absorption technique was applied to studies of autoionization in helium \cite{OttScience2013,WangPRL2010,OttNature2014}. In the so-called reconstruction of attosecond beating by interference of two-photon transitions (RABITT) technique, trains of attosecond pulses combined with IR probing allow for phase measurements, e.g. in the vicinity of bound \cite{SwobodaPRL2010} and autoionizing states \cite{HaesslerPRA2009}. Autoionizing resonances have also been studied using a single harmonic and a delayed IR probe \cite{DoughtyCP2011}, as well as by XUV pump/XUV probe spectroscopy \cite{SkantzakisPRL2010}. This topic has stimulated vigorous theoretical activity \cite{WickenhauserJMO2006,CaillatPRL2011,ArgentiPRL2010,ChuPRA2012,StrelkovPRA2014,JimenezPRL2014,SuPRL2014,DahlstromJPB2014}.

\begin{figure}
\vspace{0.2cm}
	\centering
		\includegraphics[width=\linewidth]{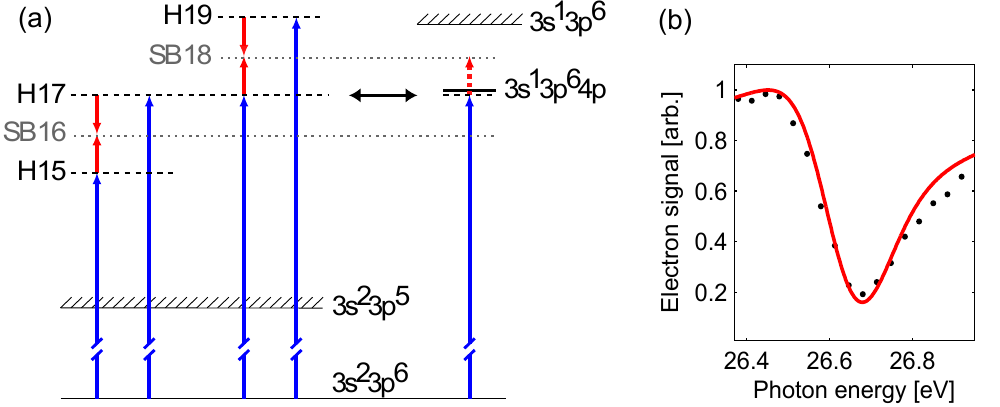}
	\caption{(a) Ar energy diagram showing the states, channels and processes involved in the present work. The blue arrows represent ionization at different harmonic frequencies. The red arrows denote absorption or stimulated emission of IR photons. One-photon ionization leads to 3s$^2$3p$^5$ $\varepsilon \ell$, $\ell$=s,d continua while two-photon ionization leads to 3s$^2$3p$^5$ $\varepsilon \ell$, $\ell$=p,f continua. The energy of harmonic 17 can be tuned across the 3s3p$^6$4p resonance, which decays by autoionization (black arrow). The processes indicated by the red dashed arrows are briefly discussed below. (b) Photoionization signal as a function of harmonic 17 energy.}
	\label{fano}
\end{figure}

In this Letter, we present an interferometric study of photoionization of argon in the proximity of the 3s$^2$3p$^6$ $\rightarrow$ 3s$^1$3p$^6$4p autoionizing resonance [Fig.~1] using 
a coherent XUV comb of odd-order harmonics of a tunable fundamental field. 
A synchronized, weak IR probe field stimulates two-photon ionization, where, in addition to absorption of an XUV photon, an IR photon is either absorbed or emitted, giving rise to sidebands in the photoelectron spectrum at energies corresponding to the absorption of an even number of IR photons. Quantum interference between the pathways involving two neighboring harmonics leads to oscillation of the sideband signals as a function of pump/probe delay \cite{PaulScience2001}. The phases of the oscillations for sidebands 16 and 18 strongly depend on the detuning of harmonic 17 from the resonance. Our experimental measurements and theoretical calculations, based on a perturbative model, \cite{JimenezPRL2014} show that the phase of an ionizing wavepacket is strongly distorted, in a non trivial way, i.e. different from a $\pi$ jump, by the presence of a quasi-bound state. We give an interpretation for the phase variation, which reflects the interaction between the continuum $3p^{-1}\varepsilon s,d$ and the quasi-bound $3s^{-1}4p$ states.
 
We used an amplified titanium sapphire laser system, which delivers 5\,mJ, 20\,fs pulses, centered around 800\,nm, at a repetition rate of 1\,kHz. Two acousto-optical programmable dispersive filters were used to achieve either a bandwidth of up to 100\,nm or a 50\,nm tunability range for the central wavelength at a reduced bandwidth of $\sim$50\,nm.
The pulses were directed into an actively stabilized Mach-Zehnder interferometer similar to that described in \cite{KroonOL2014}. In one of the arms, high-order harmonics were generated in a pulsed gas cell, followed by a 200\,nm thick aluminum foil which removed the leftover fundamental field. The harmonic and the IR pulses were collinearly recombined and focused into a diffusive gas target in the interaction region of a magnetic bottle electron spectrometer \cite{KruitJPE1983}. 

\begin{figure}
\vspace{0.2cm}
	\centering
		\includegraphics[width=\linewidth]{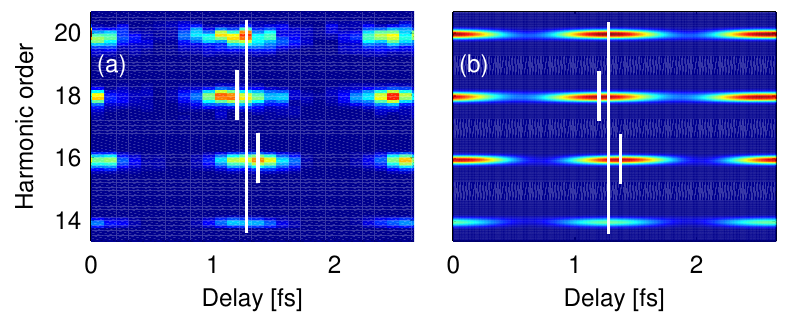}
	\caption{Sideband photoelectron spectra as a function of delay, at an excitation energy of 26.63\,eV; (a) Experimental data, corrected for the chirp of the attosecond pulses; (b) Theoretical calculations. Photoionization signal at the harmonic energies has been removed from the spectra for clarity. The short lines indicate the position of sidebands 16 and 18. The long lines join the maxima of sidebands 14 and 20.} 	
	\label{sideband}
\end{figure}

Photoelectron spectra in the kinetic energy range between 4\,eV and 20\,eV, corresponding to photoionization with harmonics 13-23, were recorded as a function of the delay between the XUV and the IR pulses. When the laser wavelength is such that the 17th harmonic is detuned to be far from the resonance, the delays corresponding the sideband maxima depend linearly on the electron energy. This dependence arises from the intrinsic chirp of the attosecond pulses \cite{MairesseScience2003}, somewhat reduced by the anomalous dispersion of the aluminum foil \cite{LopezMartensPRL2005}. The data presented in the following are corrected for the attosecond chirp, which is estimated, independently of the excitation wavelength, by linear interpolation between the maxima of sidebands 14 and 20. The intensity of the probe beam was kept as low as possible, in order to suppress processes involving absorption or emission of more than one IR photon. It also allows us to neglect the influence of the probe field on the resonance, in contrast to previous work where the probe changes the position and characteristics of the autoionizing resonance \cite{OttScience2013}.    

Figure~2(a) shows sideband oscillations after correcting for the chirp of the attosecond pulses, at an energy of harmonic 17 slightly on the blue side of the autoionizing resonance. Clearly, the maxima of sidebands 16 and 18 are shifted in opposite directions. The sideband peaks are broadened due to the XUV and IR field bandwidths, the spectrometer resolution, and the spin-orbit splitting (0.17\,eV), which is not resolved in the experiment. We also verified, by changing the gas for the generation of harmonics, that the observed shift of the sideband maxima was not a property of the harmonic radiation. Figure~2(b) shows theoretical results obtained by using the method described below. To extract the phase of the oscillation, the sideband signal was fitted to an interference equation $S=A \cos(2\omega \tau - \Delta\phi) + C$ where $\tau$ is the time delay between the XUV and the IR pulses, $\omega$ the IR frequency, $\Delta\phi$ is the phase of the oscillation, $A$ its amplitude and $C$ a constant offset.   
 
Figure~3 presents the key results of this work, with the phase variations of sidebands 16 (a) and 18 (b) as a function of the photon energy of harmonic 17. The corresponding photoionization signal due to harmonic 17 shows the characteristic  behavior of a window resonance [Fig.~1(b)]. The black symbols are the experimental results, while the red solid and green dashed lines show theoretical calculations.  Fig.~3(a,b) show a significant phase variation, by almost 0.6 radian, across the resonance. The phase variation is asymmetric, with a bias towards positive values for sideband 16 and negative values for sideband 18. 

\begin{figure}[hb]
	\centering
		\includegraphics[width=\linewidth]{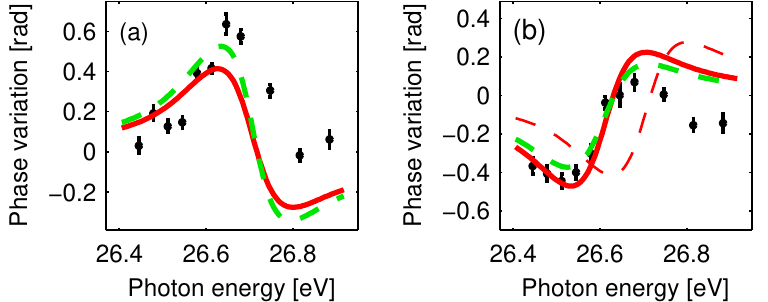}
		\caption{Phase variation of sideband 16 (a) and sideband 18 (b), as a function of the energy of harmonic 17. The theoretical results are indicated by the red solid line, while the experimental results are shown by the black symbols. The thin dashed red line in panel (b) is the opposite of the red line for sideband 16, which is close to the corresponding results for sideband 18, apart from an energy shift. The green dashed lines correspond to calculations including the processes indicated with dashed arrows in Fig.~1 (a).}
	\label{f:all_delays}
\end{figure}

Theoretical calculations were performed using a method which follows the theory developed by Fano and others \cite{FanoPR1961,StaracePRA1977} to account for the interaction between the continuum channels and the quasi-bound state and generalizes it to include the influence of a weak IR field, in the perturbative limit. Here we briefly describe the essence of this model \cite{JimenezPRL2014}, in the particular case investigated in this work. In this simplified derivation, we do not take the exchange of IR photons from the bound channel into account (as indicated by the dashed arrows in Fig.~1) and only consider the long pulse limit. The full model \cite{JimenezPRL2014}, which is compared to the experimental results, is however able to account for these effects.

The transition matrix element for two-photon ionization involving the absorption of a harmonic photon $\Omega$ and the absorption/emission of an IR photon from the ground state $g$ to a final continuum state $\psi_{\gamma E_f}$, labeled by its energy $E_f$ and angular channel $\gamma$, can be written as
\be
M_{\gamma E_f,g} = \sum_\alpha\sumint dE\,\frac{\langle \psi_{\gamma E_f}|T|\psi_{\alpha E}\rangle\langle \psi_{\alpha E}|T| g\rangle}{E_g+\Omega-E+i0^+}.
\label{eq:M2}
\ee
Here we use atomic units. An integral-sum is performed over all intermediate states $\psi_{\alpha E}$ in the discrete or continuum spectrum (energy $E$) for each of the possible channels $\alpha$. $T$ is the dipole transition operator. The multi-channel character of the transition is essential here, since for argon the $^1$P$^o$ autoionizing states that converge to the $3s3p^6$ threshold decay through two independent channels: $3s^23p^{5}\varepsilon s $ and $3s^23p^{5}\varepsilon d$. Finally $\gamma$ indicates any of the three possible final states with either S or D symmetry, $[3p^{-1} \varepsilon p]_S$, $[3p^{-1} \varepsilon p]_D$, and $[3p^{-1} \varepsilon f]_D$, whose contributions must be summed incoherently to obtain the overall sideband intensity \cite{TomaJPB2002,GuenotJPB2014}. Eq.~(\ref{eq:M2}) uses intermediate states $\psi_{\alpha E}$, which are wavefunctions of the unperturbed Hamiltonian ($H_0$). In order to include the interaction with the bound state 3s$^{-1}$4p, described by a wavefunction $|\varphi\rangle$ and with an energy $E_\varphi$, we follow the well known Fano formalism \cite{FanoPR1961} for the case of a single bound state interacting through a perturbation $V$ with two continuum channels.  
The continuum wavefunctions $|\psi_{\alpha E}\rangle$, $|\psi_{\alpha' E}\rangle$ with $\alpha$ and $\alpha'$ referring to the $s$ and $d$ continua respectively, are first transformed into interacting $|\psi_{\mathrm{1} E}\rangle$ and non-interacting $|\psi_{\mathrm{2} E}\rangle$ wavefunctions according to: 
\begin{eqnarray}
|\psi_{\mathrm{1} E}\rangle&=&\frac{V_{\alpha E}}{V_E}|\psi_{\alpha E}\rangle + \frac{V_{\alpha' E}}{V_E}|\psi_{\alpha' E}\rangle \label{eq:uni1} \\
|\psi_{\mathrm{2} E}\rangle&=&\frac{V^*_{\alpha' E}}{V_E}|\psi_{\alpha E}\rangle -\frac{V^*_{\alpha E}}{V_E}|\psi_{\alpha' E}\rangle \label{eq:uni2}
\end{eqnarray}
with $|V_E|^2= |V_{\alpha E}|^2+|V_{\alpha' E}|^2$, $V_{\alpha,\alpha'\, E}= \langle \psi_{\alpha,\alpha'\, E}|V| \varphi \rangle$. Obviously, $\langle \psi_{\mathrm{2} E}|V|\varphi\rangle=0$, while $\langle \psi_{\mathrm{1} E}|V|\varphi\rangle=V_E$. This transformation allows us to simplify the problem to that of a bound state interacting with a single continuum channel. We now diagonalize the full Hamiltonian ($H=H_0+V$) in the $\{|\varphi\rangle,\,|\psi_{1E}\rangle,\,|\psi_{2E}\rangle\}$ basis, which leads to states expressed as
\begin{equation}
|\Psi_{\mathrm{1} E}\rangle=\frac{\sin\Delta_E}{\pi V_E} |\Phi\rangle -\cos\Delta_E |\psi_{\mathrm{1} E}\rangle \,\,;\,\, |\Psi_{\mathrm{2} E} \rangle = |\psi_{\mathrm{2} E}\rangle, \\
\ee
with
\be
|\Phi \rangle=|\varphi\rangle + \emph{P} \sumint dE'\frac{V_{E'}|\psi_{\mathrm{1} E'} \rangle }{E-E'},
\label{Psi}
\ee
$\emph{P}$ denoting the Cauchy principal value. 
The quantity $\Delta_E$ is the phase shift of $\Psi_{\mathrm{1} E}$ with respect to $\psi_{\mathrm{1} E}$, 
\be
\Delta_E=-\tan^{-1} \left[ \frac{\pi |V_{E}|^2}{E-E_\Phi} \right]
\, ; \,
E_\Phi=E_\varphi+\emph{P} \sumint dE'\frac{|V_{E'}|^2}{E-E'}.
\nonumber
\ee
Introducing the parameter $q$ and the reduced energy $\epsilon$  
\be
q= \frac{\langle \Phi|T| g\rangle }{\pi V^*_E \langle \psi_{\mathrm{1}E}|T| g\rangle} \, ; \,
\epsilon= \frac{E-E_\Phi}{\pi|V_E|^2}
\label{red}
\ee
the one-photon transition matrix elements become 
\be
\langle \Psi_{\mathrm{1}E}|T| g\rangle=\langle \psi_{\mathrm{1}E}|T| g\rangle \,\frac {q+\epsilon}{\epsilon+i} \, ; \,\langle \Psi_{\mathrm{2}E}|T| g\rangle=\langle \psi_{\mathrm{2}E}|T| g\rangle. 
\label{1phot}
\ee
The one-photon ionization cross-section is the sum of the absolute squares of these matrix elements. The non-interacting channel contributes a smooth background to the Fano profile. 
The two-photon transition matrix element [Eq.~(\ref{eq:M2})] can be written, after some manipulations, as
\be
M_{\gamma E_f,g} = M^{(1)}_{\gamma E_f,g} \frac {q+\epsilon}{\epsilon+i} + M^{(2)}_{\gamma E_f,g},
\label{eq:Mq}
\ee
where $M^\mathrm{(j)}_{\gamma E_f,g}$ ($\mathrm{j}=1,2$) has a similar expression as Eq.~(\ref{eq:M2}) with the intermediate wavefunctions now equal to $|\psi_{\mathrm{j} E} \rangle$, and where $q,\epsilon$ are calculated at the energy $E_g+\Omega$. All of these quantities smoothly vary with energy and the resonant effects are contained in the ratio $(q+\epsilon)/(\epsilon +i)$.  
To determine the unitary transformation defined in Eqs.~(\ref{eq:uni1},\ref{eq:uni2}), we used complex partial transition amplitudes derived from a multi-configuration Hartree-Fock (MCHF) approach \cite{CarettePRA2013}. 
The strength of the transition amplitudes between the non-resonant components of the intermediate and final continuum channels were estimated in the plane-wave approximation~\cite{JimenezNJP2013}. To compare with the experimental results, it is necessary to account for the finite duration of the attosecond pulse train, estimated to be $\sim 12$\,fs, and of the fundamental IR field (25\,fs). In addition, a small blue shift of the generating fundamental field relative to the probe field was included, to account for ionization-induced dispersion effects in the generating medium. Figure~3 shows the results of our model as a solid red line. The agreement with the experimental data, both for the relative absorption cross-section of harmonic 17 and for the phase variation of sidebands 16 and 18, is convincing, especially on the low energy side of the resonance. 

In principle, the phase variation of sidebands 16 and 18 should be of opposite sign since the resonance affects the path corresponding to absorption of an IR photon for sideband 16, whereas it affects the path where an IR photon is emitted for sideband 18. The red dashed line in Fig.~3(b) is the opposite of the variation of sideband 16. It is quite close to the red line representing the variation of sideband 18, apart from an energy displacement, which is due to the blue shift discussed above. This effect moves the position of the resonance by an estimated $55$ meV for sideband 16 and $-55$ meV for sideband 18.
 
We also examined the influence of other possible processes induced by the interaction of the resonant state with the IR field, as represented by the dashed arrows in Fig.~1(a). These processes can be included by changing $q$ into a complex parameter $q^{\pm}_{\beta}=q \pm 2(q-i)\beta \omega/\Gamma$, with the $+$ sign for sideband 16 and the $-$ sign for sideband 18. $q^{\pm}_{\beta}$ approaches $q$ when the couplings between the intermediate, quasi-bound part of the (radial) wavefunction and the final continua are weak. 
The best agreement between the model's predictions and the experimental measurements is found when $\beta$ is very close to zero ($\beta=0.005$) as indicated by the green dashed curves in Fig.~3.

\begin{figure}
	\centering
		\includegraphics[width=\linewidth]{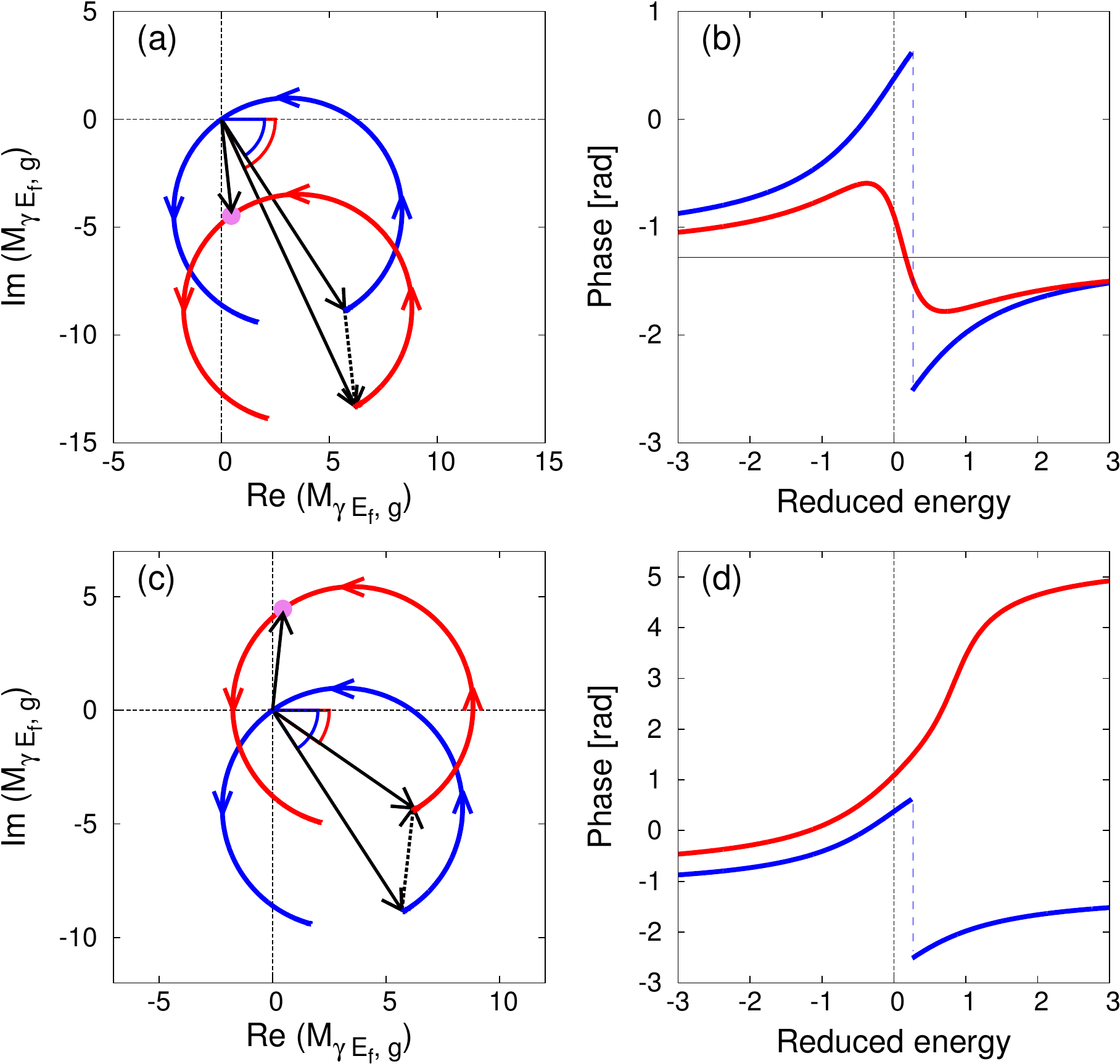}
		\caption{(a) Complex plane representation of $M_{\gamma E_f, g}$ (red circle) and of its resonant  $M^{(1)}_{\gamma E_f,g}(q+\epsilon)/(\epsilon+i)$,(blue circle) and non-resonant $M_{\gamma E_f,g}^{(2)}$ (magenta dot) components; (b) Phase variation of $M_{\gamma E_f,g}$ across the resonance ($q=-0.25$) in the absence of (blue line) and in the presence of (red line) a non-resonant component; (c) and (d) Similar representations for the opposite phase of the non-resonant contribution.}
\end{figure} 

The measured phase profile across the autoionizing resonance can be interpreted by considering the multichannel nature of the problem, explicit in the matrix element $M_{\gamma E_f,g}$ [Eq.~(\ref{eq:Mq})]. Fig.~4(a) shows the trajectory of $M_{\gamma E_f,g}$ in the complex plane, its resonant and nonresonant contributions, as the reduced energy varies from -3 to 3. The resonant contribution describes counterclockwise a circle that passes through the origin for $\epsilon=-q$. In the absence of a contribution from the non-resonant channel, the phase of $M_{\gamma E_f,g}$ follows the phase of the Fano profile, $\arg\left[(q+\epsilon)/(\epsilon+i)\right]$. 
The phase increases first steadily with energy, then drops discontinuously by $\pi$ when $\epsilon=-q$, to increase again thereafter, until it attains its original value [blue line in Fig.~4(b)]. For small values of $q$ as is the case in the present work ($q =-0.25$, \cite{SorensenPRA1994,BerrahJPB1996}), the $\pi$ phase jump occurs close to $\epsilon=0$ and the phase variation is almost symmetric.
In the presence of a non-interacting channel, the phase variation across the resonance will in general differ from $\pi$. In the present case, the non-resonant complex amplitude moves the circular trajectory away from the origin, as indicated by the phase vectors (phasors) in Fig.~4(a). As a result, the phase of the total amplitude varies smoothly across the resonance [red line in Fig.~4(b)], and by a total amount less than $\pi$. In other systems, the non-resonant amplitude could shift the circle towards the origin instead of away from it [see example in Fig.~4(c)], in such a way that the trajectory of the total amplitude encircles the origin. In that case, the total phase variation is close to $2\pi$ [Fig.~4(d)]. The dispersion of the photoelectron wavepacket is strongly affected by the resonance and quite different in the two cases. 

The phase of a two-photon ionization matrix element, with the intermediate step in the continuum \cite{DahlstromCP2013}, is the sum of the phase accumulated in the one-photon ionization step and a contribution from the continuum-continuum transition. The latter phase depends weakly on the angular momentum and is similar for the resonant and nonresonant contributions. Our numerical calculations show that the phase of the two-photon ionization matrix element indeed follows that of the one-photon ionization step, and thus carries information on the dynamics of the one-photon induced autoionization process. 

In summary, we have measured, for the first time, the distortion of the phase of the continuum induced by the coupling with an autoionizing state, using an interferometric method based on two-photon ionization. The experimental results are well reproduced by a calculation based on ab-initio parameters for the configuration interaction. These measurements could be extended to other types of Fano resonances, e.g. in He, where absorption/emission of IR photons from the localized quasi-bound state is expected to play a more important role. 

We thank Richard Squibb and Stefanos Carlstr\"om for careful reading of the manuscript. This work was partly supported by the European Research Council (Grant PALP), the Marie Curie ITN MEDEA, the Knut and Alice Wallenberg Foundation, the Swedish Research Council, and the Swedish Foundation for Strategic Research. AM, EL and MD acknowledge the support of the Kavli Institute for Theoretical Physics (National Science Foundation under Grant No. NSF PHY11-25915). AM acknowledges the support of the grant ANR-11-IDEX-0004-02.  
AJG, LA, and FM acknowledge computer time from the CCC-UAM and Marenostrum Supercomputer Centers and financial support from the European Research Council under the European Union's Seventh Framework Programme (FP7/2007-2013)/ERC grant agreement 290853, the MINECO project FIS2013-42002-R, the European COST Actions XLIC CM1204, and the Marie Curie ITN CORINF. 

\hyphenation{Post-Script Sprin-ger}


\end{document}